\documentclass[12pt]{article}
\usepackage{amsmath,amssymb,amsthm,amsfonts}

\usepackage{amsmath,amssymb,amsfonts}
\usepackage{graphicx}

\def\RR{{{\mathbb R}}}

\title{An inversion method based on random sampling for real-time MEG neuroimaging}
\author{Annalisa Pascarella\thanks{ {\it Istituto per le Applicazioni del Calcolo ``M. Picone'' - C.N.R.}, Via dei Taurini, 00161 Roma, Italy.
e-mail: {\tt a.pascarella@iac.cnr.it}},
 Francesca Pitolli\thanks{{\it Dept. SBAI, University of Roma ''La Sapienza''}, 
Via A. Scarpa 16, 00161 Roma, Italy. 
e-mail: \tt{francesca.pitolli@sbai.uniroma1.it}} 
}

\date{}

\begin{document}

\maketitle

\begin{abstract}
The MagnetoEncephaloGraphy (MEG) has gained great interest in neurorehabilitation training 
due to its high temporal resolution. The challenge is to localize the active regions of the brain
in a fast and accurate way. In this paper we use an inversion method based on
random spatial sampling to solve the real-time MEG inverse problem. 
Several numerical tests on synthetic but realistic data show that the method  
takes just a few hundredths of a second on a laptop to produce an accurate map of the electric activity inside the brain.
Moreover, it requires very little memory storage. For this reasons
the random sampling method is particularly attractive in real-time MEG applications.
\end{abstract}

\section{Introduction.}
The MagnetoEncephaloGraphy (MEG) is a completely non-invasive neuroimaging technique 
with a high  temporal resolution -~in the millisecond scale~- which can be used to 
map fast cerebral responses to spontaneous and/or evoked stimuli \cite{supek14}. For this reason MEG imaging 
has been recently used in real-time applications, such as brain-computer interface training 
or neurofeedback rehabilitation \cite{sudre2011}. To infer information on
the location of the brain active regions we need to reconstruct the 
neuroelectric current distribution underlining the magnetic data. 
This results in a highly ill-posed and ill-conditioned inverse problem \cite{KS04}.
In this paper we use a method based on the random sampling to solve the MEG inverse problem
at a low computational cost. The numerical tests show that the method takes less then $10^{-1}$ second on a laptop to produce an accurate 
neuroelectric current map.

The paper is organized as follows. The MEG inverse problem is outlined in Section~2 while 
Section~3 is devoted to the random sampling method. In Section~4 the results of several numerical tests 
on syhthetic but realistic data are shown.
Finally, some conclusion are drawn in Section~5.
\section{The MEG inverse problem.}
The MEG inverse problem consists in reconstructing the neuroelectric current  
flowing inside the brain having available a set of measurements of the neuromagnetic field generated
in the outer space by one or more neuroelectric sources.
To solve the inverse problem we first have to set up the forward model, i.e. the model relating the external magnetic field 
and the electric current distribution inside the brain. As usual, we model the brain 
as a conducting volume and assume that just small frequencies are involved in the biological phenomena we are interested in.
Under these assumptions, the neuro-electromagnetic field can be modeled by the quasi-static Maxwell's equations 
(see \cite{DPTR01,HHIKL93} and references therein for details).     
Thus, the forward integral operator $\mathbf B(\textbf{q},\textbf{J})$,
representing the magnetic field outside the head generated by 
a current distribution $\textbf{J}$ inside the conducting brain volume $V_0$, reduces to the Amp{\`e}re-Laplace law
\begin{equation} \label{AmpLap}
\mathbf B(\textbf{q},\textbf{J})= \frac{\mu_0}{4 \pi}  \int_{V_0} \, \frac{\textbf{r}'-\textbf{q}}{|\textbf{r}'-\textbf{q}|^3}
\times { \textbf{J}(\textbf{r}\,')}  \, d\textbf{r}\,'\,, \qquad \textbf{q} \notin V_0\,,
\end{equation}
where $\mu_0$ is the permeability of the brain, usually  assumed the same as the permeability of the vacuum.
\\
MEG devices sample the magnetic field using few sensors located on a helmet, placed on the head
of the subject under study. Here, we consider MEG devices equipped with magnetometers, 
which measure just the normal -~w.r.t. the skull~- component of the magnetic field. 
Let us denote by $\textbf{q}_i$, $i=1,\ldots,N$, the $N$ sites where the magnetometers are located
and by ${\mathbf e}(\textbf{q}_i)$ the normal unitary vector in  $\textbf{q}_i$.
Thus, projecting (\ref{AmpLap}) along ${\mathbf e}(\textbf{q}_i)$ we get
\begin{equation} \label{projB}
 B_i(\textbf{J})= \mathbf B(\textbf{q}_i,\textbf{J}) \cdot {\mathbf e}(\textbf{q}_i) = \frac{\mu_0}{4 \pi}  \int_{V_0} \, \left ({\mathbf e}(\textbf{q}_i)
\times \frac{\textbf{r}'-\textbf{q}_i}{|\textbf{r}'-\textbf{q}_i|^3}\right)
\cdot{ \textbf{J}(\textbf{r}\,')}  \, d\textbf{r}\,'\,.
\end{equation}
The MEG inverse problem results in minimizing the discrepancy 
\begin{equation}\label{inverse_problem}
\Delta(\textbf{J})=\sum_{i=1}^N(G_i-B_i(\textbf{J}))^2\,,
\end{equation}
w.r.t. $\textbf{J}$, once the magnetic data $G_i$, $i=1,\ldots,N$, are given.  
Since $\mathbf B(\textbf{q},\textbf{J})$ has a non-trivial kernel, additional constraints 
have to be added in order the inverse problem be feasible (see \cite{KS04} and reference therein).
\section{The random sampling method.}
In MEG applications we have just a few hundreds of magnetic data from which we want to reconstruct 
the neuroelectric current map in tens of thousands of voxels having a side of few millimeters length. 
Since the neuroelectric current distribution can be assumed to be spatially sparse, we expect that only few elementary 
sources might be sufficient to represent the unknown current \cite{Do92,FP08}.
To enforce sparsity, we model the current $\textbf{J}$ as a sum of {\em elementary sources} belonging to a large dictionary, {\em i.e.},
\begin{equation} \label{decompJ}
\mathbf J(\mathbf r) = \sum_{k} \, \mathbf J_k \, 
\psi_k(\mathbf r)\,, 
\end{equation}
where $\mathbf J_k = (J_k^x,J_k^y,J_k^z)$ is the electric current vector of the $k$ elementary source 
having a {\em ``small''} spatial distribution $\psi_k$. 
\\
Assuming that $\mathbf J(\mathbf r)$ can be compressed by the basis $(\psi_k)$,
so that just few elementary sources are sufficient to 
well reconstruct the unknown quantity $\mathbf J(\mathbf r)$, 
in the random sampling method \cite{BPP14,PP14} we select randomly few sources in the dictionary,
{\em i.e.},
\begin{equation} \label{decompJM}
\mathbf J(\mathbf r) \approx \sum_{k\in {\cal K}} \, \mathbf J_k \, \psi_k(\mathbf r)\,, 
\end{equation}
where ${\cal K}$ is a {\em small} subset of random indexes having cardinality $M$.
\\
Now, the discrete inverse problem consists in determining a configuration of the current density vector 
$J=(\textbf{J}_1,\ldots,\textbf{J}_M)^T$ that minimizes the discrepancy 
\begin{equation} \label{disc_inverse_problem}
\Delta(J)=\|{BJ-G}\|_{\RR^N}^2\,,
\end{equation}
where $B \in \RR^{N\times (3M)}$ is the {\em lead field matrix} with entries 
\begin{equation} \label{leadfield}
\begin{array}{l}
\displaystyle B_{ik}^{l} = \frac{\mu_0}{4 \pi} \int_{V_0} \, \left(\mathbf e(\textbf{q}_i)
\times \frac{\textbf{r}_k-\textbf{q}_i}{|\textbf{r}_k-\textbf{q}_i|^3} \right)_l \, 
\psi_k(\textbf{r}\,')   \, d\textbf{r}\,'\,, \\ \\
i=1,\ldots,N,\qquad  k=1,\ldots,M, \qquad l=x,y,z\,,
\end{array}
\end{equation} 
and $G=(G_1,\ldots,G_N)^T$ is the given measurement vector.
\\
In the random sampling method $M$ can be chosen in the order of the number of data $N$, 
so that the ill-conditioning of the inverse problem is reduced and the minimization 
of the discrepancy become feasible.
Thus, the inverse problem (\ref{disc_inverse_problem}) can be solved 
by the least squares method, without the need for additional constraints. This means that the random sampling can be seen
as a regularization method that promotes current distributions that are spatially sparse. 
Further constraints are needed in the case when we are interested in reconstructing the neuroelectric activity produced by deep sources
for the least squares method suffers from a drift toward the surface of the brain when localizing deep sources.
In this case we can use a beamforming method that consists in constructing a spatial filter that favors isolated neuroelectric activity
located in selected region of the brain \cite{SN08}. Beamforming methods are also suitable to reconstruct the activity generated by multiple sources.
\\
Finally, we notice that a few repeated runs, each one with a different sample of elementary sources, can be performed to improve the accuracy
of the reconstructed current map.
\section{Numerical tests.}
\begin{figure}[htp]
\centering
\begin{tabular}{cc}
\includegraphics[width=6cm]{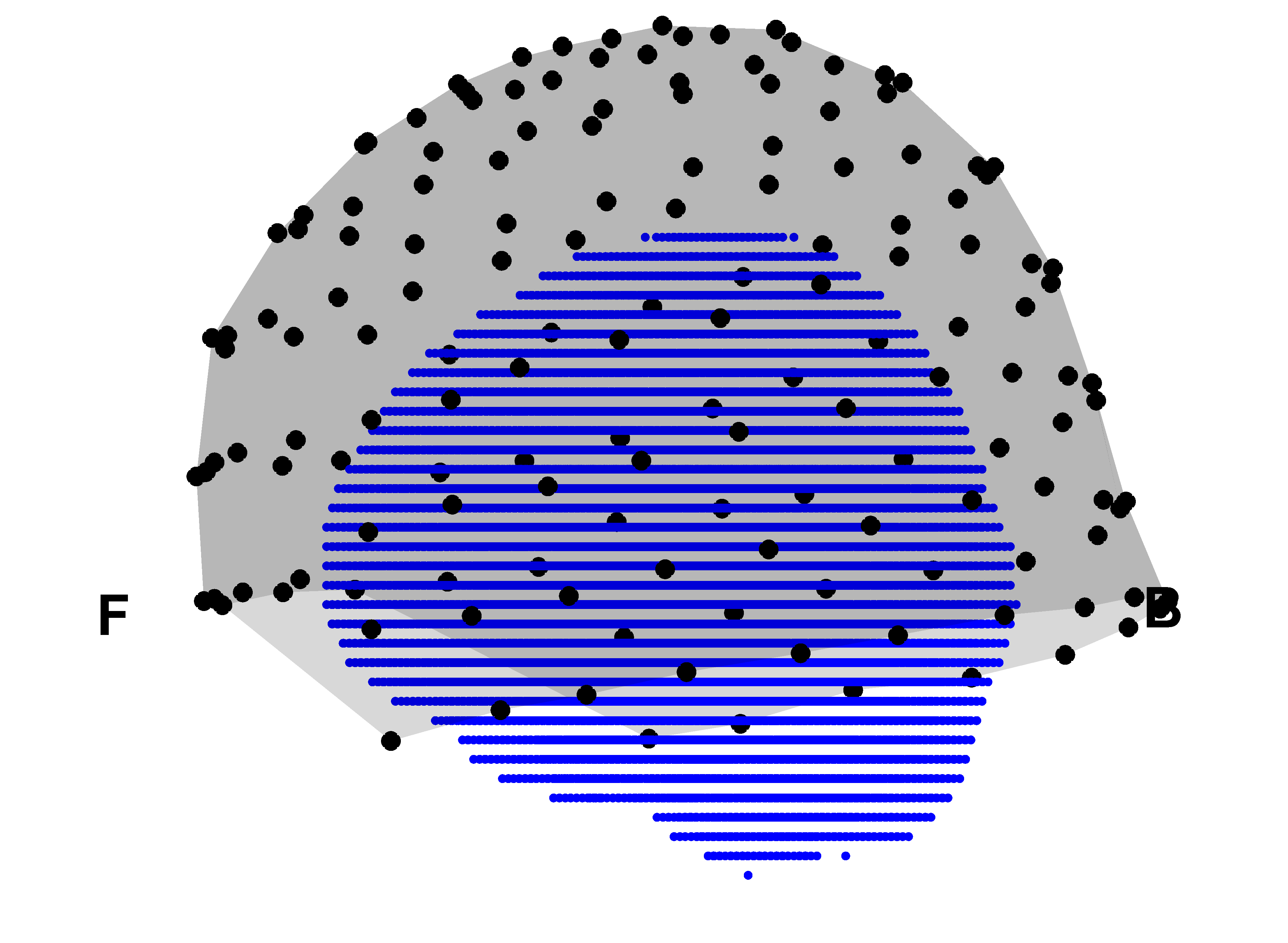}%
&
\includegraphics[width=6cm]{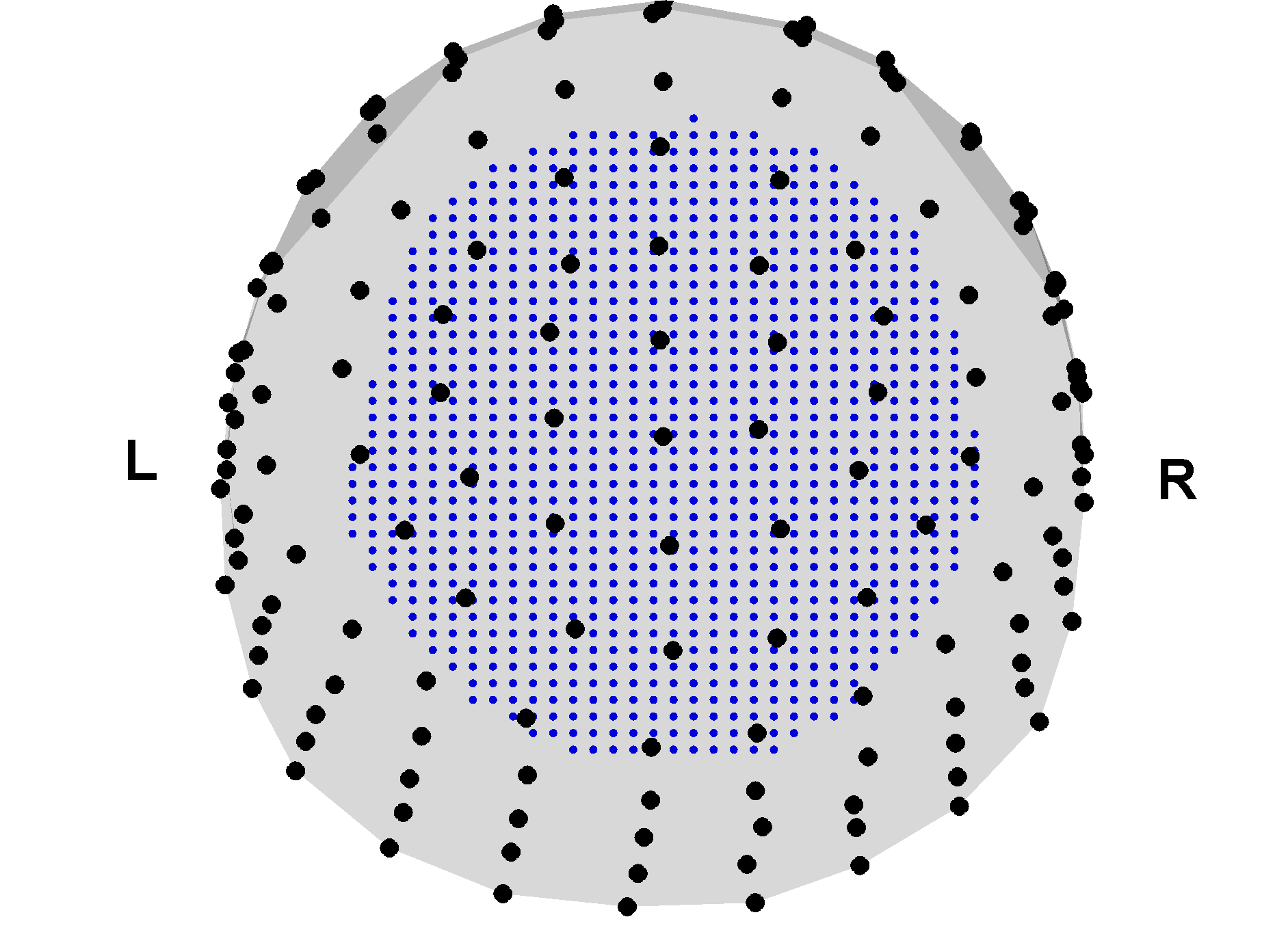}%
\end{tabular}
\caption{The source space (blue points) and the sensor sites (black circles). Left: lateral view (F: front; B: back). Right: axial view (L: left; R: right).}
\label{Setup}
\end{figure}

We tested the random sampling method on synthetic data designed in order to reproduce a realistic MEG experiments. To this end we modeled the head and the brain by
the phantom constructed by the Montreal Neurological Institute (MNI) \cite{MNI}. In particular, the MNI source space model we used is formed by 20173 points inside 
the brain volume distributed on a regular grid of 5mm edge. The head model is available in the open source software FieldTrip \cite{FT2010}. 
The sensor helmet has $153$ magnetometers located according to the MEG device used 
at Institute for Advanced Biomedical Technologies (ITAB), University  {\em G. d'Annuzio} of Chieti-Pescara 
 \cite{MEG-ITAB} (see Figure~\ref{Setup}).
Just the normal component of the magnetic field was sampled.

To generate the synthetic magnetic data we calculated the lead field matrix by the single-shell model  
available in FieldTrip (see \cite{No03} for details) using a finer source space formed by 37163 points uniformly distributed on a regular grid of 4mm edge. Then, we activated one or more current dipoles located in regions of interest (ROI) that are significant in MEG applications. 
The selected ROIs and dipole sources are listed in Table~\ref{Sources} where also the depth, that is the minimum distance between the dipole and the sensors, is reported. 
The ROIs are classified according to the AAL Atlas \cite{AAL2002}. 
\begin{table}[htp]
\caption{Source current dipole location.}
{\begin{tabular}{c|l|c}
Test & ROI & Depth (mm) \\
\hline
1 & Superior Frontal Gyrus (SFG)   & 58   \\     
2 & Superior Frontal Gyrus (SFG)   & 75   \\     
3 & Supplementary Motor Area (SMA) & 62   \\     
4 & Supplementary Motor Area (SMA) & 90   \\     
5 & Superior Parietal Gyrus (SPG)  & 61   \\     
6 & Superior Parietal Gyrus (SPG)  & 77   \\     
\hline
\end{tabular}}
\label{Sources}
\end{table}

In the following tests we assume the elementary source $\psi_k$ is a point-like source, i.e.
\begin{equation} \label{point-like}
\psi_k(\mathbf r) = \delta(\mathbf r-\mathbf r_k)\,,
\end{equation}
where $\delta$ is the Dirac function.  As a consequence, the entries of the lead field matrix in (\ref{leadfield}) reduce to 
\begin{equation}
B_{ik}^{l} = \frac{\mu_0}{4 \pi} \frac{1}{M}\left(\mathbf e(\textbf{q}_i)
\times \frac{\textbf{r}_k-\textbf{q}_i}{|\textbf{r}_k-\textbf{q}_i|^3}\right)_l\,.
\end{equation} 
Then, we select randomly the
elementary sources by selecting a sample ensemble of few random points, $\textbf{r}_k$, $k=1, \ldots M$, 
uniformly distributed inside the brain volume $V_0$.

\subsection{Single source}
\label{Test_single}
\begin{figure}[!t]
\centering
\begin{tabular}{cc}
\includegraphics[width=6cm]{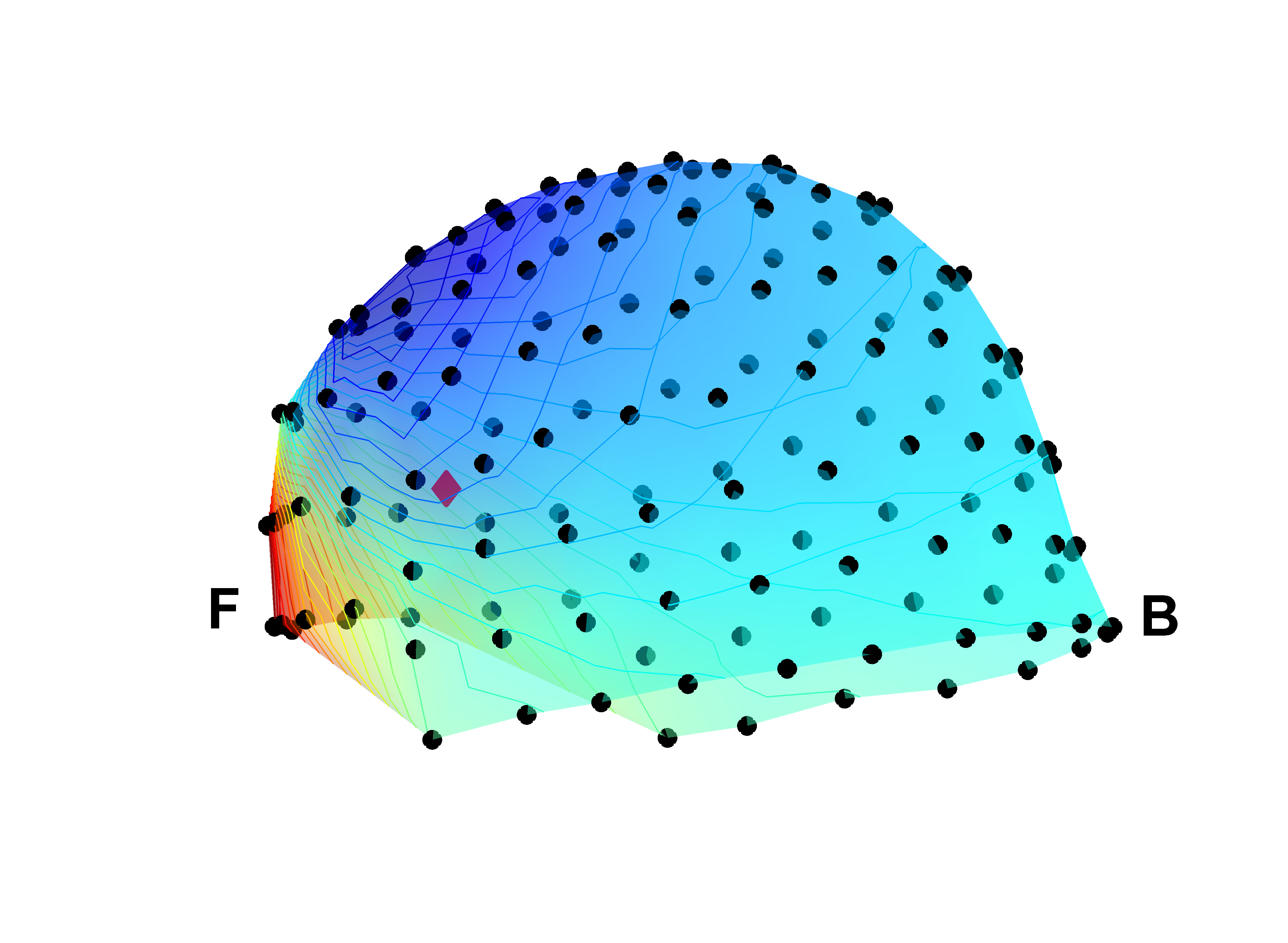}%
&
\includegraphics[width=6cm]{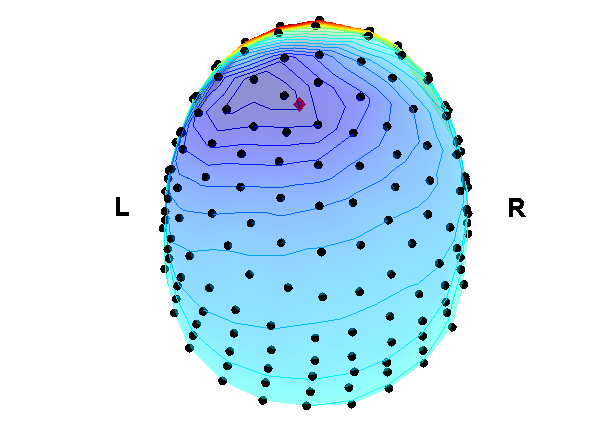}%
\end{tabular}
\caption{Test 1: The contour plot of the synthetic magnetic field generated by a superficial current dipole (red diamond)  
located in the Superior Frontal Gyrus of the brain. Left: lateral view  (F: front; B: back). Right: axial view (L: left; R: right). 
The sensor sites are displayed as black points.}
\label{Campo_SFGs}
\end{figure}

\begin{figure}[!t]
\centering
\begin{tabular}{cc}
\includegraphics[width=6cm]{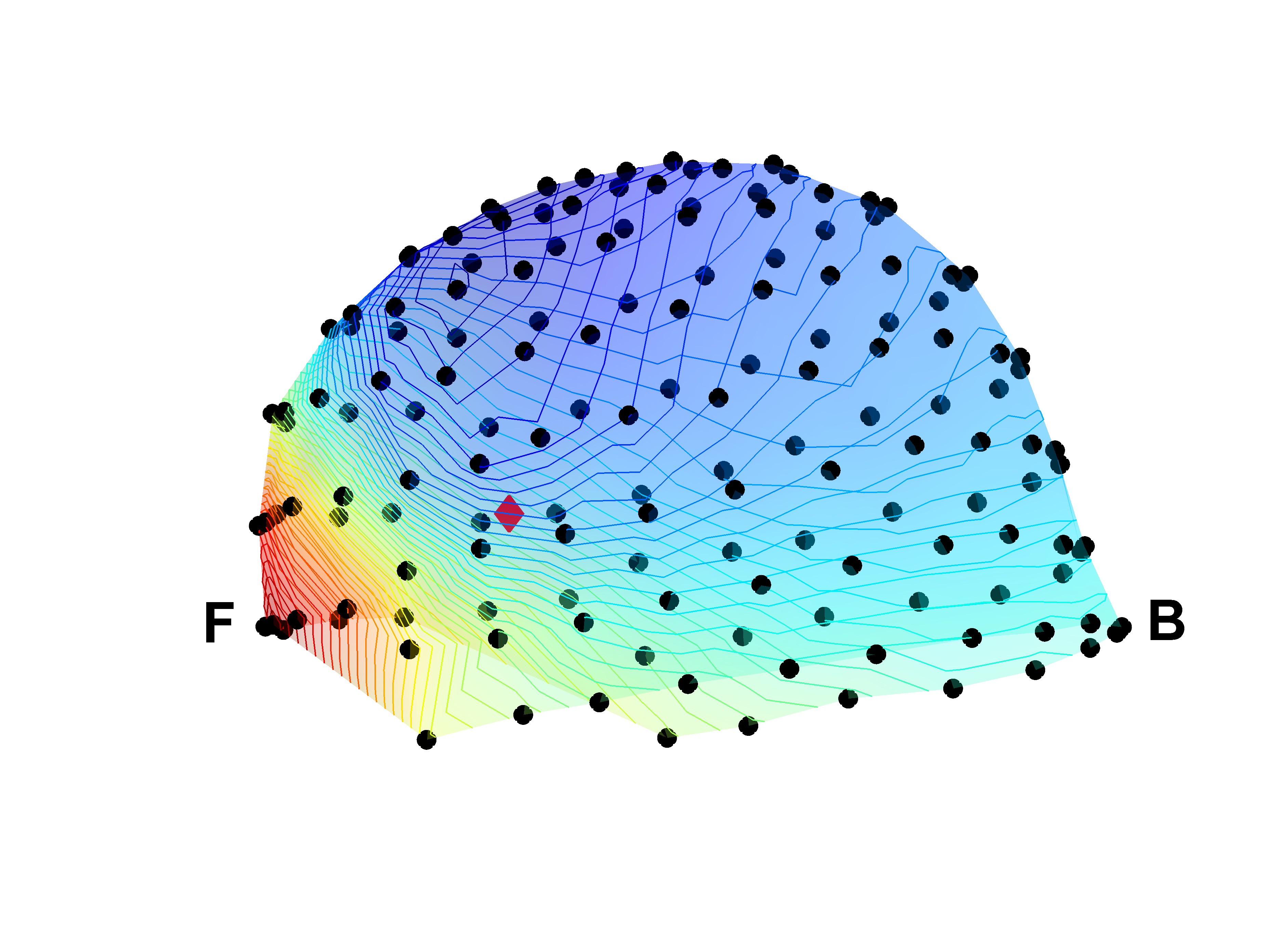}%
&
\includegraphics[width=6cm]{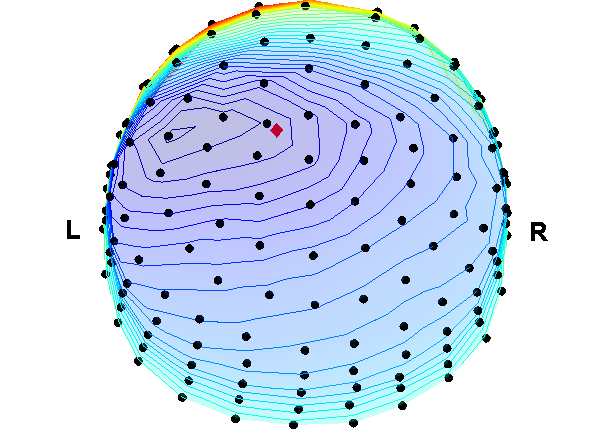}%
\end{tabular}
\caption{Test 2: The contour plot of the synthetic magnetic field generated by a deep current dipole (red diamond)  
located in the Superior Frontal Gyrus of the brain. Left: lateral view  (F: front; B: back). Right: axial view (L: left; R: right). 
The sensor sites are displayed as black points.}
\label{Campo_SFGd}
\end{figure}

First of all we tested the random sampling method in the case when the synthetic data are generated by a single current dipole.  
As an example, the synthetic data used for Test~1 and Test~2 are displayed in Figure~\ref{Campo_SFGs} and Figure~\ref{Campo_SFGd},
respectively.

\begin{figure}[!t]
\centering
\begin{tabular}{cc}
\includegraphics[width=6cm]{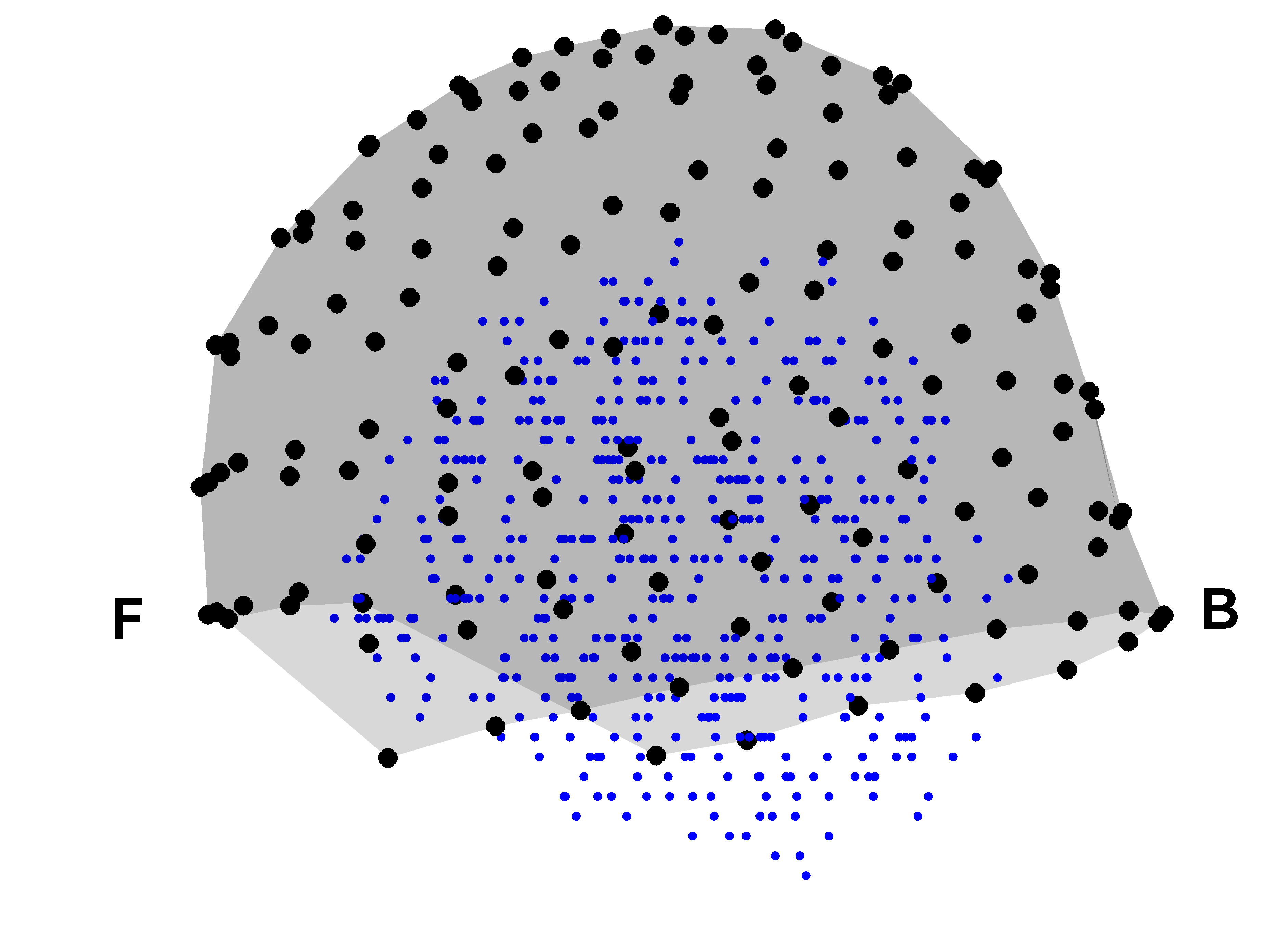}%
&
\includegraphics[width=6cm]{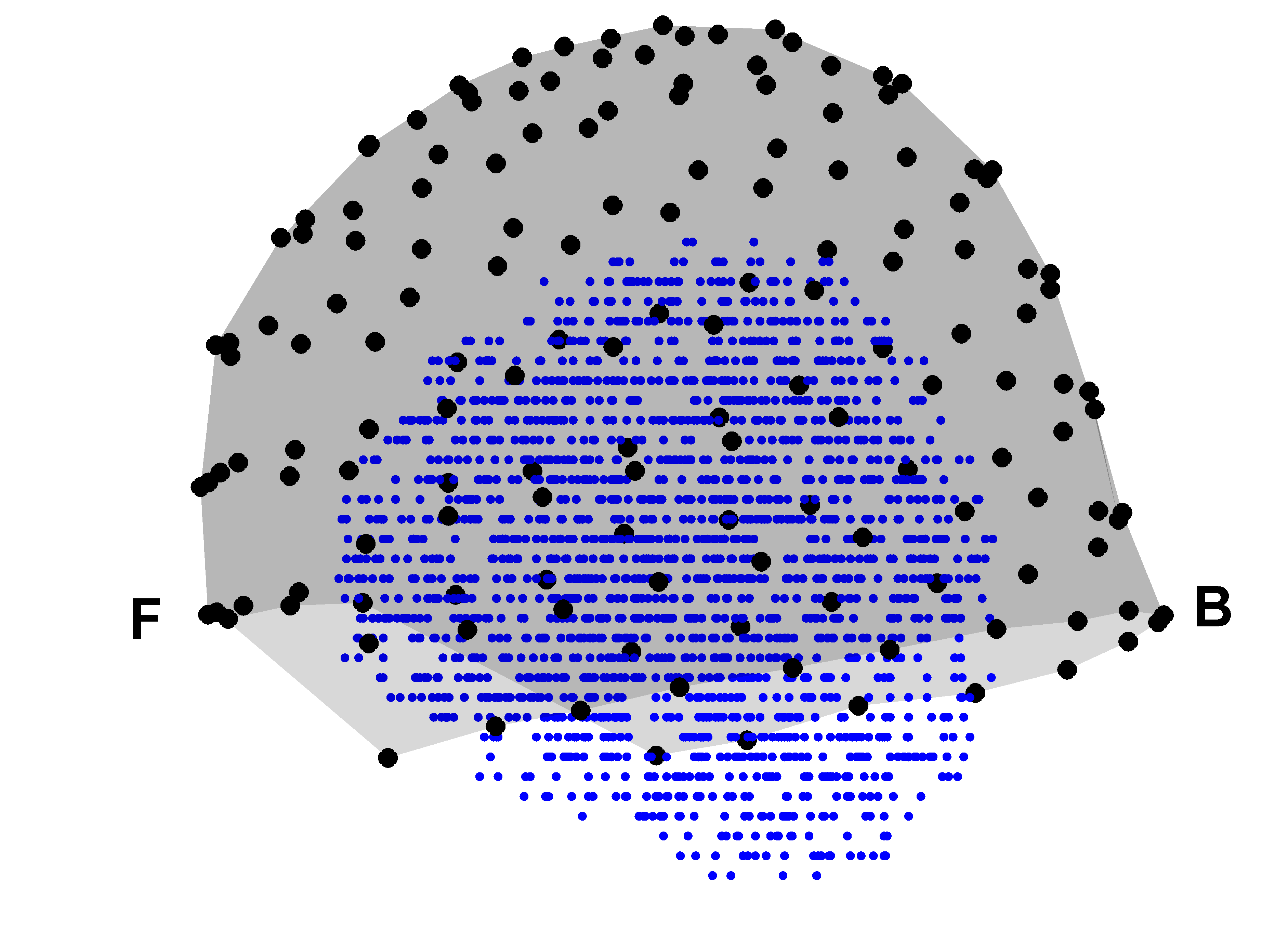}%
\end{tabular}
\caption{Two samples of random points  (F: front; B: back). Left: $M=500$. Right: $M=2000$.}
\label{RandomPoints}
\end{figure}

The electric current density inside the brain was reconstructed by the random sampling method using point-like elementary sources centered in $M$ random points 
uniformly sampled in the MNI source space. In the tests we used $M=500,1000,2000$. Two samples of random points are displayed in Figure~\ref{RandomPoints}.

\begin{figure}[!t]
\centering
\begin{tabular}{cc}
\includegraphics[width=5.5cm]{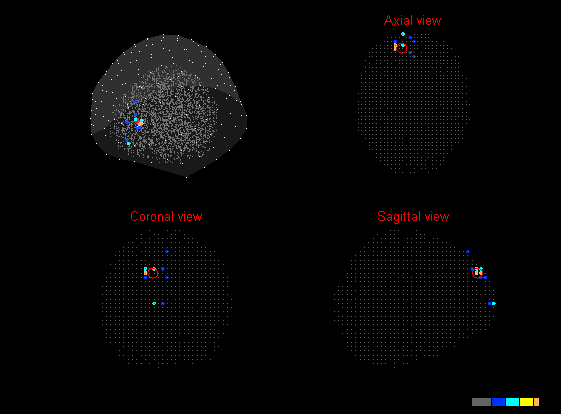}%
&
\includegraphics[width=5.5cm]{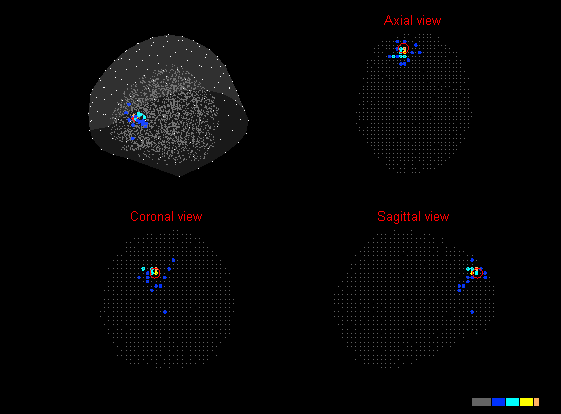}%
\end{tabular}
\caption{Test 1: The intensity of the reconstructed current using the least squares method (left) and the beamforming method (right) with $M=500$ and 5 repeated independent runs for
a total of 2500 points in the source space. The points are colored according to the intensity (from gray to orange). 
The current dipole generating the data is displayed as a red circle. The sensor sites are displayed as white points.}
\label{Intensity_SFGs}
\end{figure}

\begin{figure}[!t]
\centering
\begin{tabular}{cc}
\includegraphics[width=5.5cm]{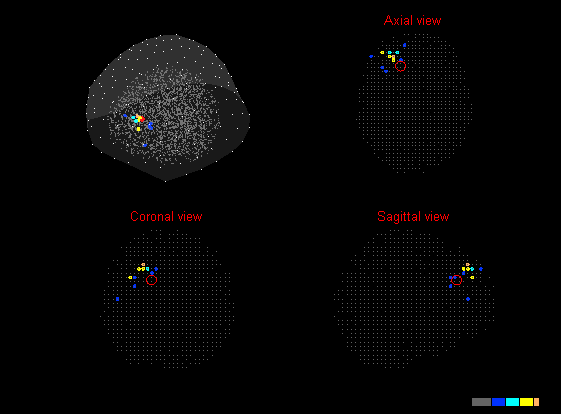}%
&
\includegraphics[width=5.5cm]{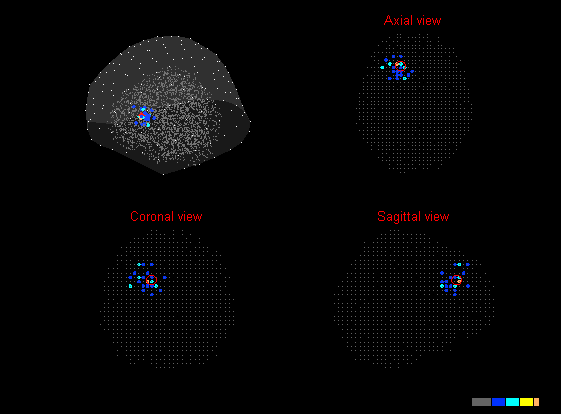}%
\end{tabular}
\caption{Test 2: The intensity of the reconstructed current using the least squares method (left) and the beamforming method (right) with $M=500$ and 5 repeated independent runs for
a total of 2500 points in the source space. The points are colored according to the intensity (from gray to orange). 
The current dipole generating the data is displayed as a red circle. The sensor sites are displayed as white points.}
\label{Intensity_SFGd}
\end{figure}

The inverse problem was solved both by the least squares method and by the beamforming method introduced in \cite{CWDS11}.
The numerical solution takes few hundredths of seconds on a laptop.
\\
To increase the accuracy of the reconstruction we repeated the run from 5 to 50 times,
each time using a different sample of random points. Then, we collected together the reconstructed current distribution obtained in each run.  
As an example, the intensity of the reconstructed current obtained after 5 repeated runs with $M=500$ is shown in Figure~\ref{Intensity_SFGs}
for Test~1 and in Figure~\ref{Intensity_SFGd} for Test 2.

\begin{figure}[!t]
\centering
\begin{tabular}{cc}
\includegraphics[width=6cm]{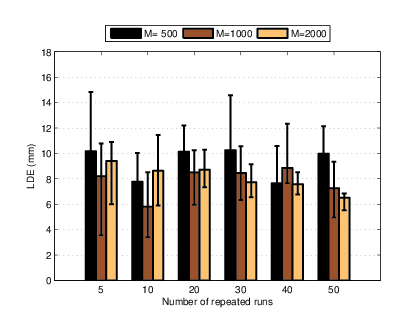}%
&
\includegraphics[width=6cm]{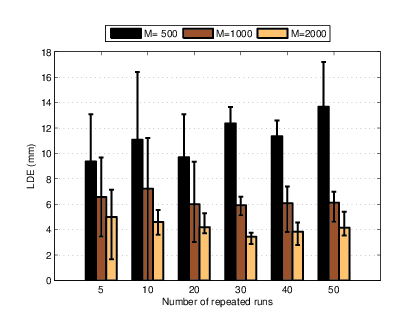}%
\end{tabular}
\caption{Test~1: The Localization Distance Error (LDE) for different number of repeated runs. Different color bars refers to different numbers of random points. 
Left panel refers to the least squares method while right panel refers to the beamforming method.}
\label{LDE_SFGs}
\end{figure}

\begin{figure}[!t]
\centering
\begin{tabular}{cc}
\includegraphics[width=6cm]{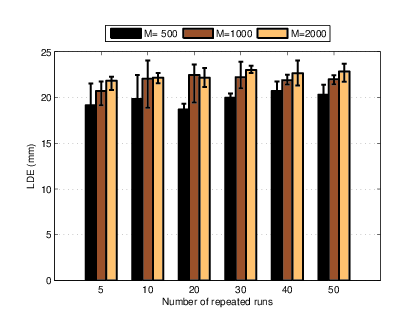}%
&
\includegraphics[width=6cm]{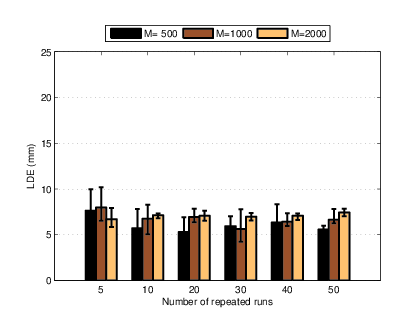}%
\end{tabular}
\caption{Test~2: The Localization Distance Error (LDE) for different number of repeated runs. Different color bars refers to different numbers of random points. 
Left panel refers to the least squares method while right panel refers to the beamforming method.}
\label{LDE_SFGd}
\end{figure}

\begin{figure}[!t]
\centering
\begin{tabular}{cc}
\includegraphics[width=6cm]{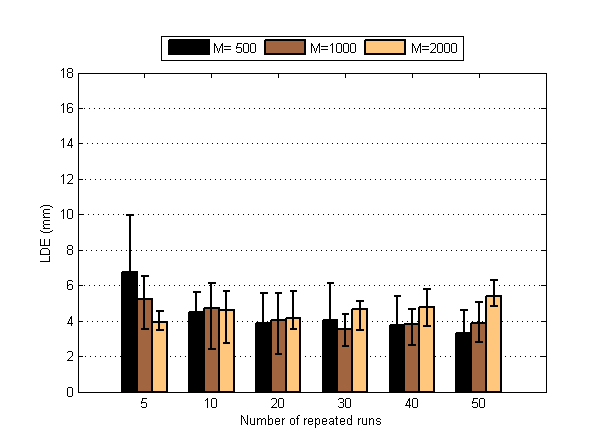}%
&
\includegraphics[width=6cm]{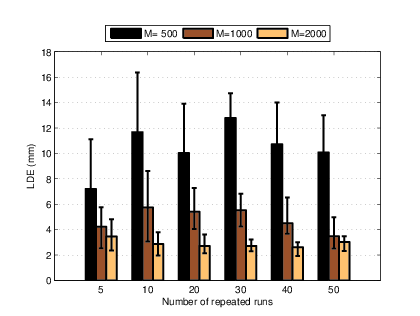}%
\end{tabular}
\caption{Test~3: The Localization Distance Error (LDE) for different number of repeated runs. Different color bars refers to different numbers of random points. 
Left panel refers to the least squares method while right panel refers to the beamforming method.}
\label{LDE_SMAs}
\end{figure}

\begin{figure}[!t]
\centering
\begin{tabular}{cc}
\includegraphics[width=6cm]{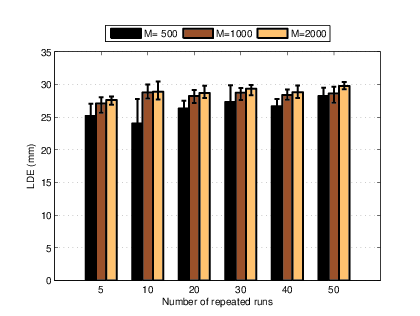}%
&
\includegraphics[width=6cm]{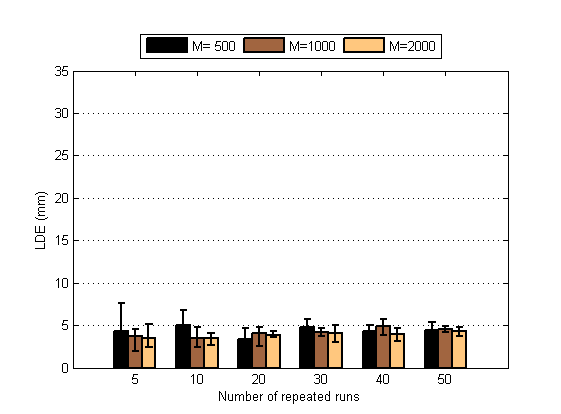}%
\end{tabular}
\caption{Test~4: The Localization Distance Error (LDE) for different number of repeated runs. Different color bars refers to different numbers of random points. 
Left panel refers to the least squares method while right panel refers to the beamforming method.}
\label{LDE_SMAd}
\end{figure}

\begin{figure}[!t]
\centering
\begin{tabular}{cc}
\includegraphics[width=6cm]{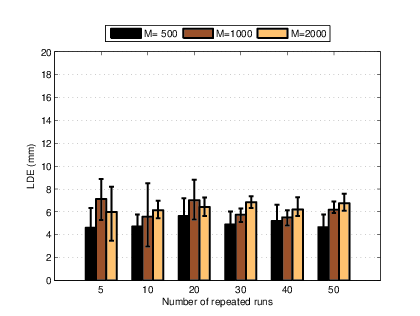}%
&
\includegraphics[width=6cm]{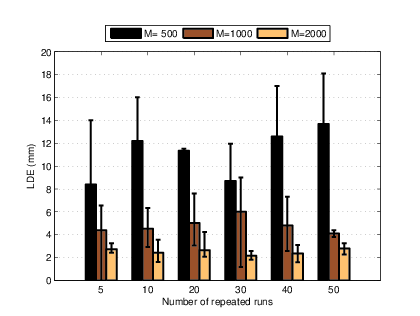}%
\end{tabular}
\caption{Test~5: The Localization Distance Error (LDE) for different number of repeated runs. Different color bars refers to different numbers of random points. 
Left panel refers to the least squares method while right panel refers to the beamforming method.}
\label{LDE_SPGs}
\end{figure}

\begin{figure}[!t]
\centering
\begin{tabular}{cc}
\includegraphics[width=6cm]{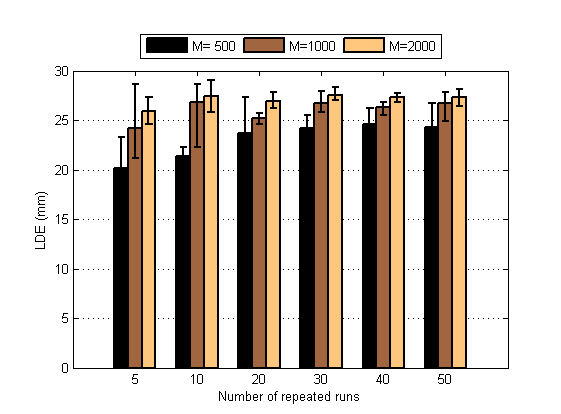}%
&
\includegraphics[width=6cm]{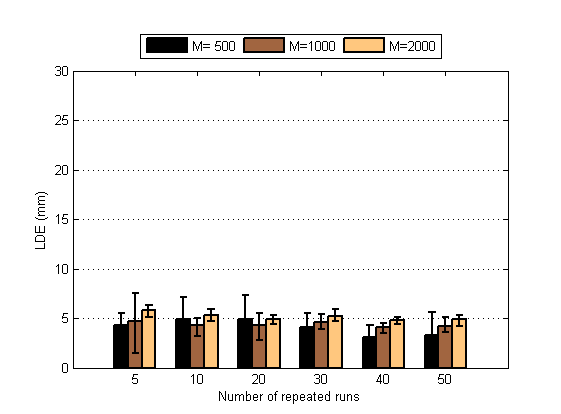}%
\end{tabular}
\caption{Test~6: The Localization Distance Error (LDE) for different number of repeated runs. Different color bars refers to different numbers of random points. 
Left panel refers to the least squares method while right panel refers to the beamforming method.}
\label{LDE_SPGd}
\end{figure}

The numerical results are summarized in Figures~\ref{LDE_SFGs}--\ref{LDE_SPGd} where the Localization Distance Error (LDE)
is shown for different number of random points -~$M=500, 1000, 2000$~- and different number of runs -~5, 10, 20, 30, 40, 50. 
The LDE is defined as the distance between the source and the center of mass of the reconstructed current.
For each value of $M$ and for each number of runs we evaluated the mean, the maximum and the minimum of the LDE.  
In the graphs the mean is represented as a bar while the error bar represents the maximum and the minimum.

\subsection{Noisy data}
\begin{figure}[!t]
\centering
\begin{tabular}{cc}
\includegraphics[width=5.5cm]{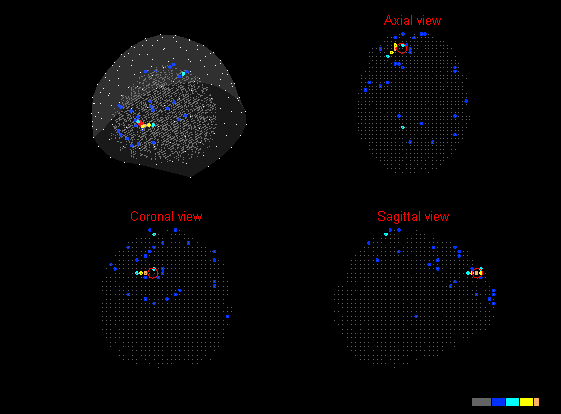}%
&
\includegraphics[width=5.5cm]{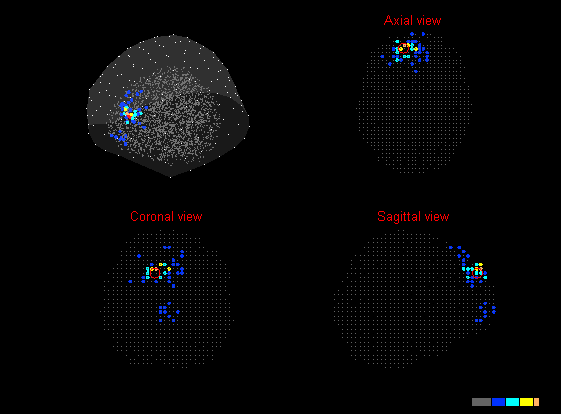}%
\end{tabular}
\caption{Test 1 (noisy data): The intensity of the reconstructed current using the least squares method (left) and the beamforming method (right) with $M=500$ and 5 repeated independent runs for
a total of 2500 points in the source space. The points are colored according to the intensity (from gray to orange). 
The current dipole generating the data is displayed as a red circle. The sensor sites are displayed as white points.}
\label{Intensity_SFGs_noise}
\end{figure}

\begin{figure}[!t]
\centering
\begin{tabular}{cc}
\includegraphics[width=5.5cm]{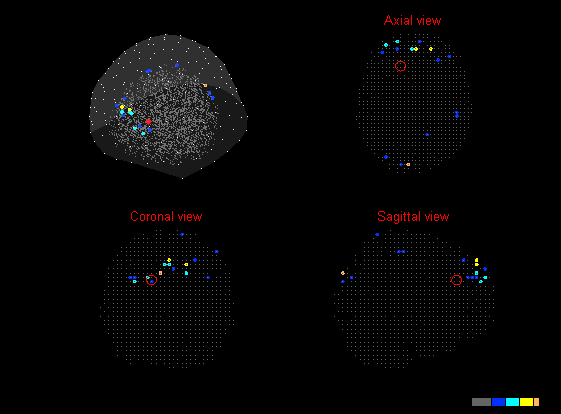}%
&
\includegraphics[width=5.5cm]{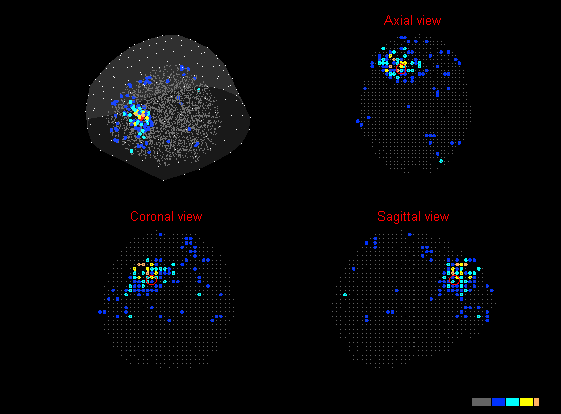}%
\end{tabular}
\caption{Test 2 (noisy data): The intensity of the reconstructed current using the least squares method (left) and the beamforming method (right) with $M=500$ and 5 repeated independent runs for
a total of 2500 points in the source space. The points are colored according to the intensity (from gray to orange). 
The current dipole generating the data is displayed as a red circle. The sensor sites are displayed as white points.}
\label{Intensity_SFGd_noise}
\end{figure}

In the second set of tests we used noisy data. The set-up of the numerical tests is the same as in Section~4.1 but
this time white Gaussian noise with $snr=25$~db was added to the synthetic data.
In Figure~\ref{Intensity_SFGs_noise} and Figure~\ref{Intensity_SFGd_noise} the intensity of the reconstructed current obtained after 5 repeated runs with $M=500$ is shown
for Test~1 and Test~2, respectively.

\subsection{Multiple sources}

\begin{figure}[!t]
\centering
\begin{tabular}{cc}
\includegraphics[width=5.5cm]{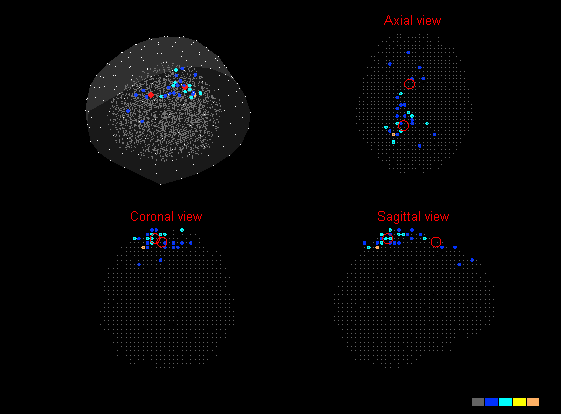}%
&
\includegraphics[width=5.5cm]{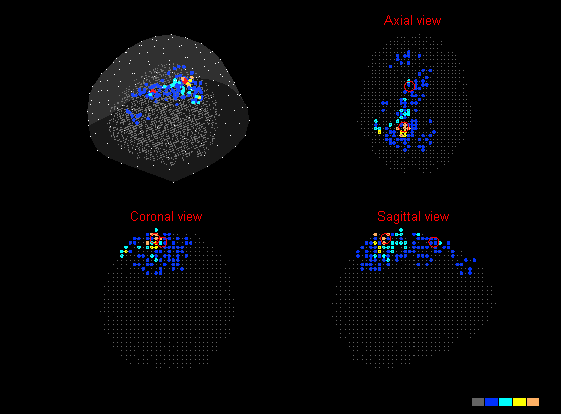}%
\end{tabular}
\caption{Test on multiple sources: The intensity of the reconstructed current using the least squares method (left) and the beamforming method (right) with $M=500$ and 5 repeated independent runs for
a total of 2500 points in the source space. The points are colored according to the intensity (from gray to orange). 
The current dipoles generating the data are displayed as red circles. The sensor sites are displayed as white points.}
\label{Intensity_two_sources}
\end{figure}

Finally, we tested the random sampling method in the case when the synthetic data are generated by two current dipoles.
In particular, the two sources are the dipole in the Supplementary Motor Area of Test~3 and the dipole in the Superior
Parietal Gyrus of Test~5. 
In Figure~\ref{Intensity_two_sources} the intensity of the reconstructed current obtained after 5 repeated runs with $M=500$ is shown.

\section{Conclusion}
The pictures show that the random sampling method allows us to produce an accurate map of the current intensity 
inside the brain at a very low computational cost, {\em i.e.}, low memory storage and low computational time. 
The method, possibly combined with the beamforming method, can easily deal
with various configuration of the neural sources, {\em i.e.}, superficial, deep or multiple sources, and produces
a rather accurate neuroelectric map also in the case of noisy data. 
\\
The bar graphs in Figures~\ref{LDE_SFGs}--\ref{LDE_SPGd} show that 
5 different runs with 5 different sample of cardinality $M=500$ in each run are sufficient to obtain a localization error
of few millimeters while increasing the number of runs and/or the number of random points in each run does not increase significantly the accuracy.
Moreover, the method does not require heavy methods, such as the boundary element method 
or the finite different method,  to evaluate the lead field matrix and can be easily adapted to different head geometry.
\\
All these reasons make the random sampling method particularly attractive in real time applications and can be possibly 
implemented on a small computer or on a tablet.

\section*{Acknowledgements.}
We would like to thank Daniela Calvetti and Erkki Somersalo for useful discussion and suggestion and for having provided us with
the algorithm to classify and display the current intensity.

\end{document}